\documentclass{nature}

\def\sro{Sr$_2$RuO$_4$}
\def\vhs{van Hove singularity}
\def\o{$^{17}$O}

\title{Pronounced drop of $^{17}$O NMR Knight shift in superconducting state of Sr$_2$RuO$_4$}

\author{A. Pustogow$^{1}$\footnote[1]{These authors contributed equally to this work.}, Yongkang Luo$^{1,2*}$, A. Chronister$^{1}$, Y.-S. Su$^{1}$, D. Sokolov$^{3}$, F. Jerzembeck$^3$, A. P. Mackenzie$^{3}$, C. W. Hicks$^3$, N. Kikugawa$^{4}$, S. Raghu$^5$, E. D. Bauer$^6$, and S. E. Brown$^1$}

\usepackage{graphicx,color,ulem}
\makeatletter
\let\saved@includegraphics\includegraphics
\AtBeginDocument{\let\includegraphics\saved@includegraphics}
\renewenvironment*{figure}{\@float{figure}}{\end@float}
\usepackage[section]{placeins}
\makeatother

\begin{document}

\maketitle

\begin{affiliations}
 \item Department of Physics $\&$ Astronomy, UCLA, Los Angeles, CA 90095, USA;
 \item Wuhan National High Magnetic Field Center and School of Physics, Huazhong University of Science and Technology, Wuhan 430074, China;
 \item Max Planck Institute for Chemical Physics of Solids, Dresden 01187, Germany;
 \item National Institute for Materials Science, Tsukuba, Japan.
 \item Geballe Laboratory for Advanced Materials, Stanford University, Stanford, CA, 94305, USA
 \item Los Alamos National Laboratory, Los Alamos, New Mexico 87545, USA.
\end{affiliations}


\begin{abstract}

The superconducting state in the quasi-two-dimensional and strongly correlated \sro\ is uniquely held up as a solid state analog to superfluid $^3$He-$A$, with an odd-parity order parameter that also breaks time reversal symmetry, and for which the vector order parameter has the same direction in spin space for all electron momenta. The recent discovery that uniaxial pressure causes a steep rise and maximum in transition temperature ($T_c$) in strained samples motivated the study of $^{17}$O nuclear magnetic resonance (NMR) that we describe in this article. A reduction of Knight shifts $K$ was observed for all strain values and temperatures $T<T_c$, consistent with a drop in spin polarization in the superconducting state. In unstrained samples, our results are in contradiction with a body of previous NMR work, and with the most prominent previous proposals for the order parameter of \sro. Possible alternative scenarios are discussed.


\end{abstract}

Following the discovery of superconductivity in the layered compound \sro, notable similarities of its low-temperature properties to the classic Fermi liquid and superfluid $^3$He were recognized~\cite{Rice1995,Mackenzie-RMP2003}. 
The normal state is a three-band material~\cite{Oguchi-PRB1995,Mackenzie-PRL1996}, with pronounced strong-correlation characteristics linked to Hund's Rule coupling of the partially filled Ru $t_{2g}$ orbitals dominating the Fermi surface. The transition to a superconducting ground state at $T_c =$1.44 K~\cite{Maeno-Nature1994}, with indirect evidence for proximity to ferromagnetism, led to the suggestion that the pair wave functions of the superconducting state likely exhibit a symmetric spin part, \textit{i.e.}, triplet~\cite{Rice1995}. Crucial support for the existence of a triplet order parameter rested on NMR spectroscopy, which showed no change in Knight shift between normal and superconducting states~\cite{Ishida-Nature1998,Duffy-PRL2000}. Later, several experiments produced evidence for time reversal symmetry breaking (TRSB)~\cite{Luke-Nature1998,Xia-PRL2006}. Together, these reports aligned well to the exciting proposal that \sro\ is a very clean, quasi two-dimensional solid-state analog of the topologically nontrivial $^3$He-$A$ phase~\cite{Leggett-RMP1975}, but here in the form of a charged superfluid.

However, several experimental results are either in direct contradiction with, or difficult to interpret in terms of a $p$-wave superconducting state.  For instance, due to the breaking of time-reversal symmetry, there is the generic expectation of measurable chiral edge currents (which propagate within a coherence length of the edge, but which are screened over a somewhat larger penetration depth scale).  However, such currents have not been detected despite several attempts with progressively improved sensitivity \cite{Bjoernsson2005,Hicks2010,Watson2018}.  Furthermore, recent thermal conductivity measurements are most naturally explained by a $d$-wave order parameter with vertical line nodes~\cite{Hassinger-PRX2017}.  Adding to the confusion, recent specific heat experiments were interpreted as evidence for a horizontal line node at $k_z=0$ , which is incompatible with an in-plane $p$-wave order parameter \cite{Kittaka2018}.

Recently, it has been shown that uniaxial pressure provides a new means of investigating the mysteries of the superconducting state of \sro. In mean-field theory, a chiral $p$-wave superconductor exhibits a split transition in the presence of in-plane uniaxial strain with an accompanying cusp in T$_c$ in the limit of zero strain.  However, strained samples of \sro \ do not exhibit  a split transition or a cusp; instead, there is a dramatic increase in T$_c$ from 1.5 K to 3.5 K, with the maximum reached at compressive strain $\varepsilon_{aa}$ estimated at -0.5\% ($\equiv\varepsilon_v$) \cite{Steppke-Science2017}. The observation was interpreted as a consequence of the Fermi energy $E_F$ of one of the three partially filled bands (the quasi-2D $\gamma$ band) meeting the Brillouin Zone boundary. Tuning through this van Hove singularity (vHs) is associated with a weak divergence (logarithmic for a strictly 2D system) in the density of states (DOS) at $E_F$, followed by a Lifshitz transition. These observations strongly motivated a study of $^{17}$O nuclear magnetic resonance (NMR) in uniaxially pressured samples; indeed, evidence for the DOS maximum came from normal state \o\ NMR experiments \cite{Luo2018}, which also demonstrated an enhanced Stoner factor $S$ accompanying the DOS peak, and suggested close proximity to a quantum phase transition to a ferromagnetic state. The results have bearing on the chiral $p$-wave state hypothesis: on the one hand, an odd-parity order parameter vanishes at the location of the vHs, thereby reducing the impact of the DOS enhancement; on the other hand, the enhanced Stoner factor and ferromagnetic fluctuations could strengthen the pairing instability.

The focus of this paper is \o\ NMR Knight shift studies on uniaxially pressured \sro, comparing the shifts seen in the normal and superconducting states. The experiments were carried out in a variable-strain device~\cite{Razorbill}, and cover the full range of $T_c$ from 1.5 K to 3.5 K. \o\ NMR spectroscopy is directly sensitive to the spin polarization $M_s$; the expectation for an $s$-wave (singlet) superconductor is a reduction in the (spin) paramagnetic shift, which would vanish in the limit $T/T_c\to0$ and $B/B_{c2}\to0$, whereas for the widely proposed equal-spin-paired E$_u$ state (Table~\ref{tab:OP}),$K_s\equiv M_s(B)/B$~\cite{ShiftDef} remains unchanged from the normal state value. Summarizing the findings, onset of superconductivity leads to a substantial drop in $M_s$ for all strains measured; the zero-strain results are therefore in disagreement with those previously reported \cite{Ishida-Nature1998}.  We describe a series of tests which, we believe, explain this discrepancy. While $M_s$ remains nonzero for $T\to0$, note that quasiparticle creation is accompanied by spin polarization, which occurs for several possible reasons in applied fields $B_0\ne0$. No evidence for a change in ground state symmetry is observed as the strain is varied over the interval $\varepsilon_{aa}=[0,\varepsilon_v]$. Concluding remarks comment on implications for the nature of superconductivity in \sro.

The crystal structure of \sro\ is identical to the undoped parent compound of the ``214'' cuprates, La$_2$CuO$_4$ (Supplemental Information). As for the 214 case, the states at $E_F$ are predominantly of $d$ character; here it derives from Ru $t_{2g}$-O $\pi$ hybridization. Fig.~S1a 
depicts the orbitals dominating the $\gamma$ band, associated with the Ru, O(1), O(1$^{\prime}$) sites. The O(2) sites are in the apical positions, situated symmetrically above and below the Ru site. The device for applying uniaxial (mostly compressive) stress along the tetragonal $\mathbf{a}$-axis is shown adjacent, in Fig.~S1b. Throughout this report, the magnetic field $\mathbf{B_0}\parallel\mathbf{b}$, since out of plane field components suppress $B_{c2}$. 
On stressing the sample, the relevant response is the resulting asymmetric strain $\varepsilon_{aa}-\varepsilon_{bb}$, written below simply as $\varepsilon_{aa}$.

\begin{figure}[h]
\centering
\includegraphics[width=0.8\columnwidth]{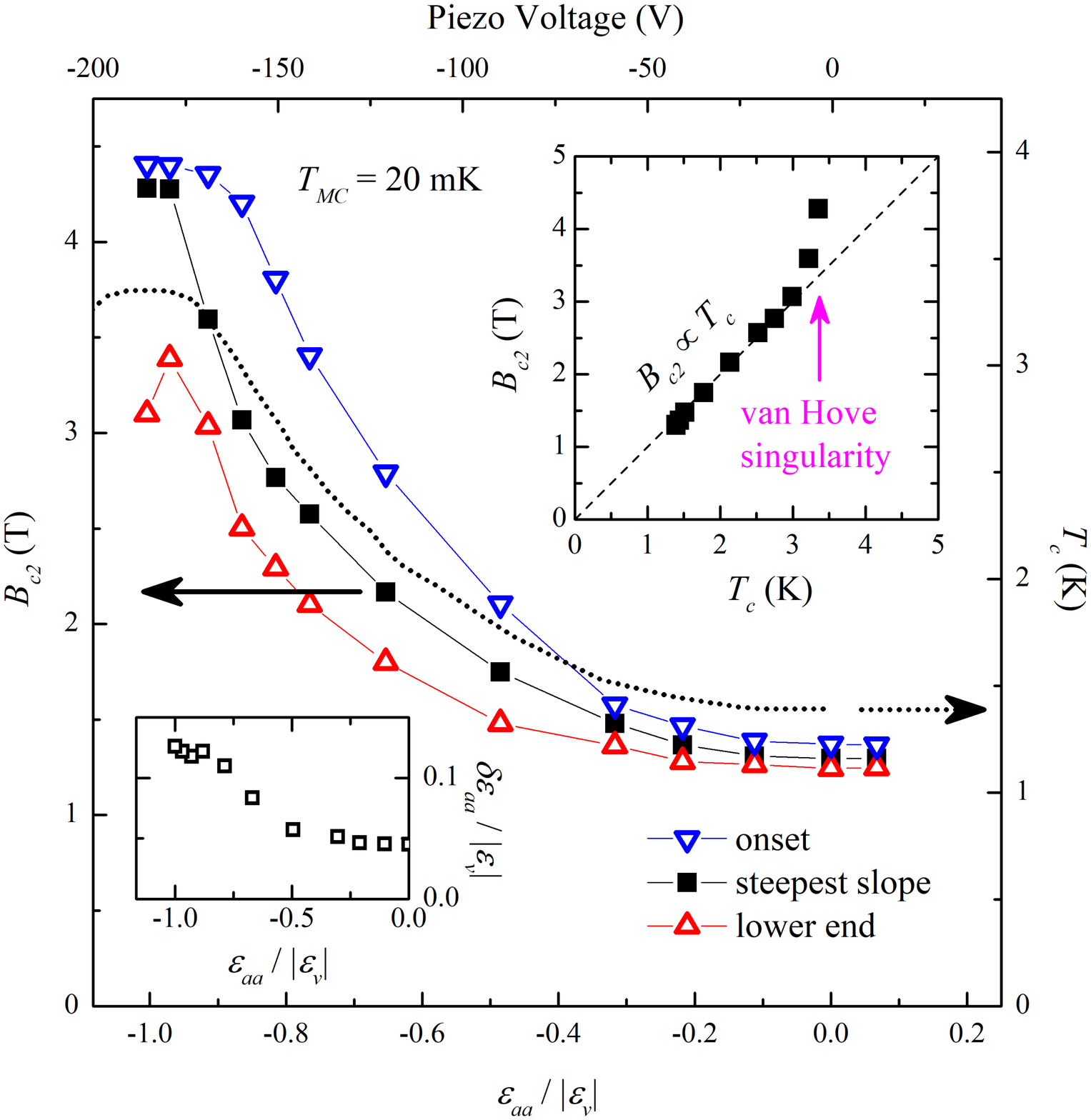}
\caption{\textbf{$|$ Strain dependence of the upper critical field of \sro.} $B_{c2}(T\to0,\varepsilon_{aa})$, determined by ac susceptibility measurements at base temperature, $T = 20$~mK. The incease with compressive strain peaks at $\varepsilon_v$, thus closely following the trend of the critical temperature $T_c$~\cite{Steppke-Science2017}. Upper inset: A linear relation $B_{c2}\propto T_c$ holds over a broad range of strain, except near to $\varepsilon_v$. Lower inset: Strain gradients become more pronounced at higher strain reaching about $0.1\varepsilon_v$ at the \vhs. More details are provided in the SI.}
\label{Hc2}
\end{figure}

Since magnetic fields lead to quasiparticle spin polarization, the ideal experiment has the applied field $B_0\ll B_{c2}$. Nuclear spin polarization, on the other hand, favors the largest field possible. For guidance in making this compromise in the choice of experimental parameters, $B_{c2}(\varepsilon_{aa})$ was first determined from the field-dependent RF power reflected from the tuned/matched tank circuit (see Supplementary Information), the results are shown in Fig. \ref{Hc2}. $B_{c2}$ is maximized at $\varepsilon_v$, coincident with $T_c^{max}$, at a value ($4.3\pm0.05$ T) that is within a few per cent of that (4.5 T) reported in Ref. \cite{Steppke-Science2017}. The reduction could be a result of a small misalignment from the in-plane condition, \textit{O}($1^\circ$)~\cite{Kittaka2009}. The minimum value is $B_{c2}(\varepsilon_{aa})=1.32\pm0.05\mathrm{T}$, identified by extending the measurements to tensile strains $\varepsilon_{aa}>0$. The upper inset shows that for much of the range studied, $B_{c2}\sim T_c$, except for very close to the vHs, where $B_{c2}$ increases more sharply relative to $T_c$.

\begin{figure}
\centering
\includegraphics[width=0.6\columnwidth]{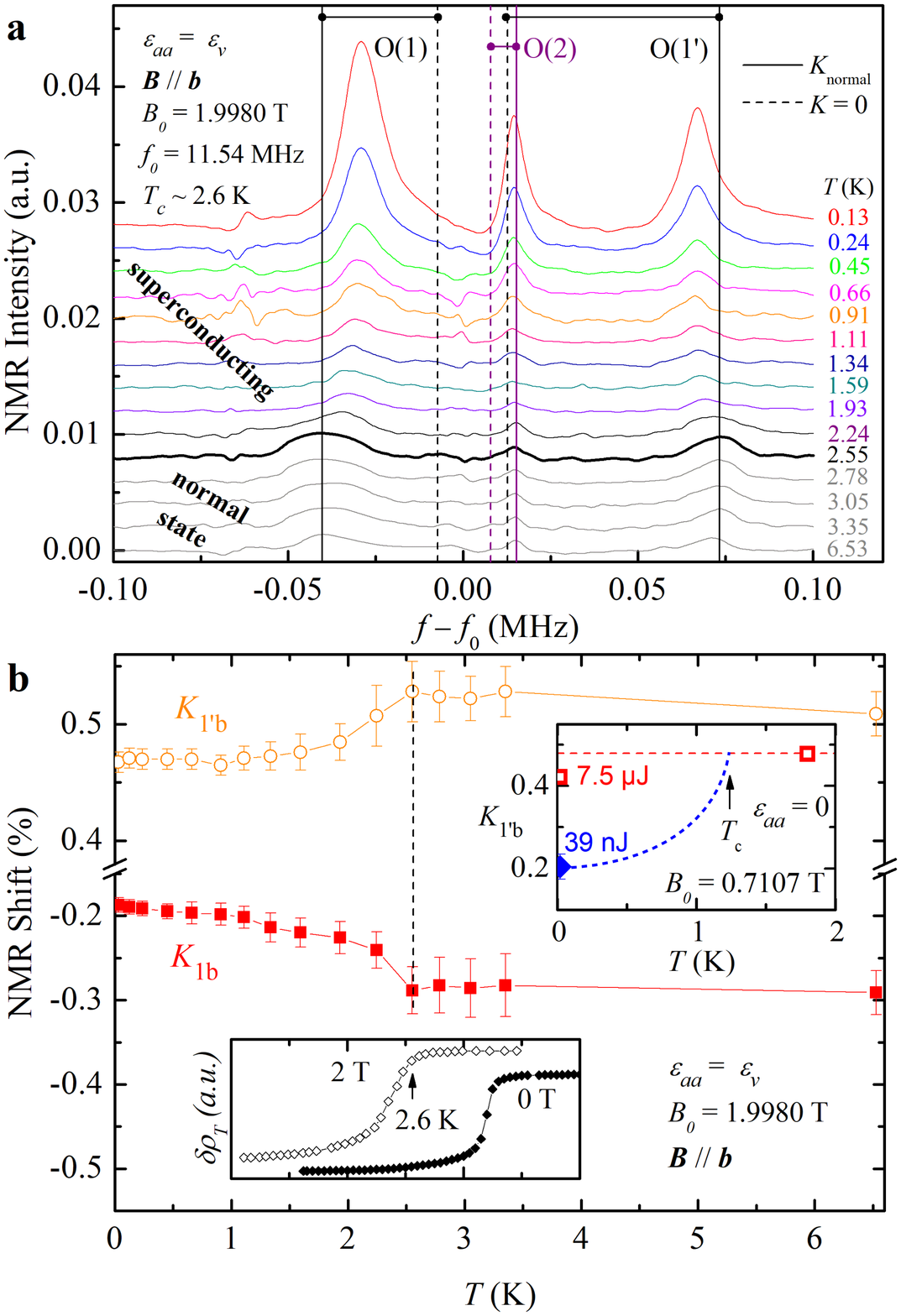}
\caption{\textbf{$|$ Knight shifts $K$ vs. $T$, measured at the \vhs\ ($\varepsilon_{aa}=\varepsilon_v$).} \textbf{a,} The NMR spectra at applied field $B_0=1.998$~T and carrier $f_0$=11.54 MHz. Shown are three peaks corresponding to the O(1), O(2) and O(1$^{\prime}$) sites (from left to right). \textbf{b,} The associated Knight shifts $K_{1b}$ and $K_{1^{\prime}b}$ show a pronounced reduction below $T_c(B_0) = 2.6$~K (see lower inset and SI), evidencing a drop of spin polarization $M_s$ in the superconducting state. Upper inset: Experiments with varied pulse energy reveal a similar decrease of $M_s$ below $T_c$ for $\varepsilon=0$; details below, see Fig.~\ref{fig:pulse-energy-FID}.}
\label{vHs-temperature}
\end{figure}

The temperature dependence of the \o\ central transition frequencies for the three sites were measured at $\varepsilon_{aa}=\varepsilon_v$, where $B_{c2}$ is largest. The RF carrier frequency is $f_0=11.54$~MHz, and the applied magnetic field $B_0=1.9980$ T. Results are shown in Fig.~\ref{vHs-temperature}, in which pronounced changes in shift for all three sites are clearly observed upon decreasing the temperature through $T_c(B_0)$. Marked also are the \textit{estimated} frequencies corresponding to zero shift for each site. Since the orbital shifts are relatively small, the (3) different ``unshifted'' references are attributed to quadrupolar effects on the central transition~\cite{Luo2018}.

Since the normal state Knight shifts $K_{1b}<0$, $K_{1'b}>0$, the changes for $T<T_c$ in Fig.~\ref{vHs-temperature}b correspond to a drop in $M_s$ of order 20--30~\%, and present a qualitatively different result from the published zero-strain results \cite{Ishida-Nature1998}. Note that the shifts $K\sim M_s/B_0$, remain nonzero for $T\to0$, where field-induced quasiparticles are likely relevant at the relatively high fields ($B_0/B_{c2}\simeq0.45$) at which the measurement was performed. In addition to the contributions from vortex cores, two other sources should be considered in the context of gap nodes, or at least the deep minima; these are the Volovik Effect \cite{Volovik1993}, and Zeeman coupling.

The observed drop of $M_s$ upon entering the superconducting state under strained conditions invites a comparison to the known results at zero strain, for which the lack of reported decrease is a cornerstone of the case for a chiral $p$-wave order parameter.  Therefore, we carried out measurements covering the entire range of strain conditions, $\varepsilon_{aa}=[0,\varepsilon_v]$. In doing so, the results were found to depend on NMR pulsing details. With this important observation in mind, a reexamination of the shifts for $\varepsilon_{aa}=0$ is presented first.

\begin{figure}
\centering
\includegraphics[width=0.8\columnwidth]{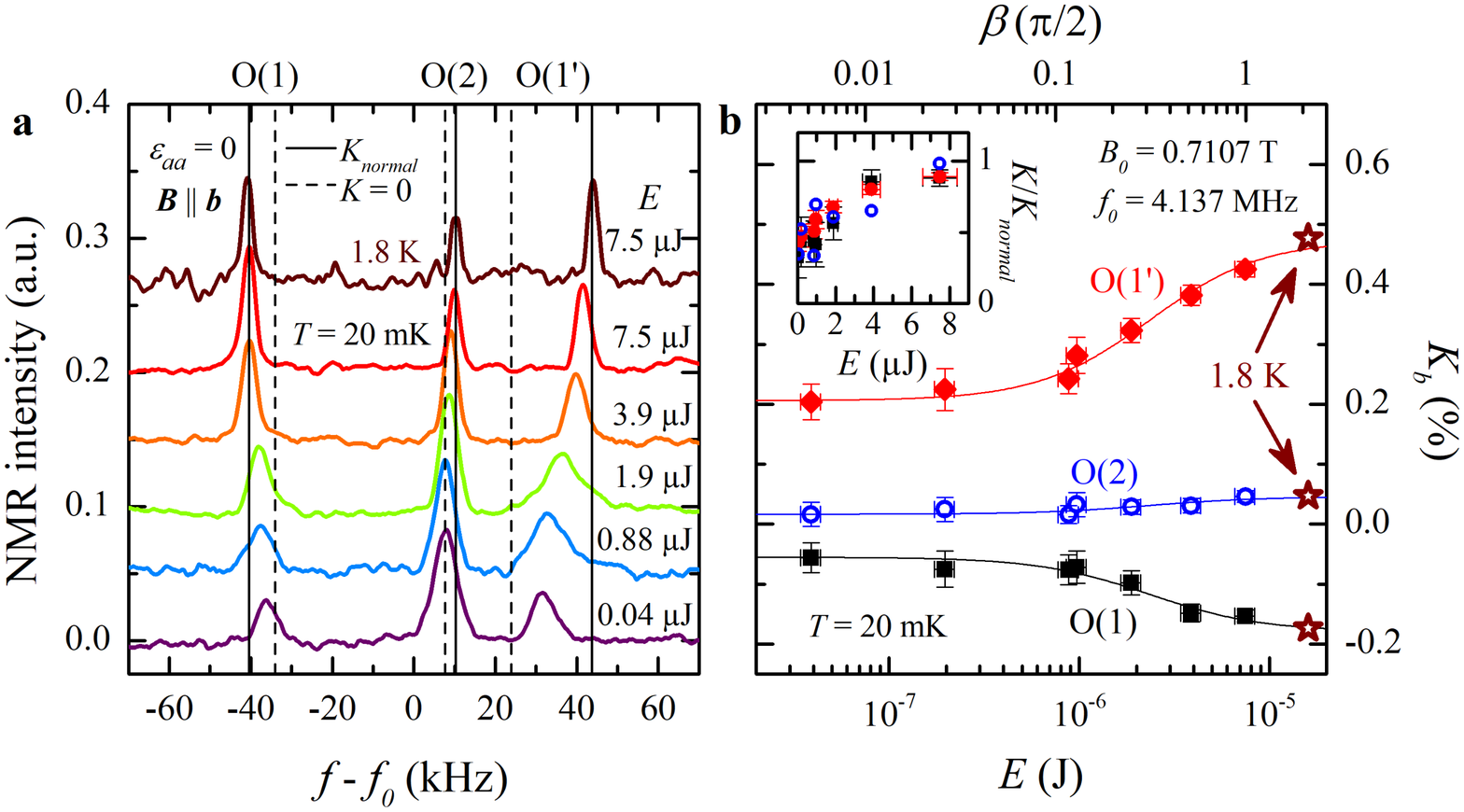}
\caption{\textbf{$|$ Zero-strain $^{17}$O NMR spectra of \sro\ for varying pulse energy.}
\textbf{a}, Free induction decay (FID) measurements with varying pulse lengths $d_1\le d_{\pi/2}$ ($\pi/2$ corresponds to $E=7.5$~$\mu$J) were applied at the nominal base temperature $T_{MC}=20$~mK, with $B_0=0.7107$ T, $f_0=4.137$ MHz. The O(1) and O(1$^{\prime}$) peak shifts indicate smaller $M_s$ for smaller $E$.  For each site, the normal-state position is marked by the solid vertical lines; the estimated $K=0$ position is the dashed line. \textbf{b}, Energy (equivalently: tip angle $\beta$) dependence of O(1), O(1$^{\prime}$) shifts. Inset: The relative Knight shift reductions (normalized to normal state). Note that the three sites give comparable reductions; uncertainties for O(2) site are not shown.}
\label{fig:pulse-energy-FID}
\end{figure}

In Fig.~\ref{fig:pulse-energy-FID} we present the spectra with no applied stress, collected following various pulse excitations. The applied field is $B_0=0.7107$~T, referenced to the $^3$He nuclear resonance condition for RF carrier frequency $f_0=4.137$ MHz, and similar to the 0.65 T field used in Ref. [\cite{Ishida-Nature1998}]. The mixing chamber temperature is $T_{MC}=20$ mK, confirmed by a measurement of the $^{63}$Cu $T_1$ (in the coil), and exploiting the accepted value $T_1T=1.27$ s-K. Note that from Fig.~\ref{Hc2}, $B_0/B_{c2}(\varepsilon_{aa}=0)\simeq0.55$. As above, the three spectral lines shown (\textit{left}$\to$\textit{right}) are the central transitions for O(1), O(2) and O(1$'$), from low to high frequency, respectively. The top trace of Fig.~\ref{fig:pulse-energy-FID}a corresponds to the normal state at $T=1.8$ K, collected using a standard two-pulse echo pulse sequence, $[\pi/2-t_1-\pi-acquire]$. The remaining spectra are all recorded at 20 mK base temperature, and transformed from signals produced following single-pulse excitations of variable time durations $d_1$, chosen because it is far less constraining than the echo sequence in regard to the amount of energy transmitted. These are ordered bottom-to-top with increasing pulse energy. The vertical dashed lines correspond to the estimated resonance position for vanishing Knight shift for each of the three sites. The inset depicts the shifts \textit{vs.} pulse energy $E$; the variations are approximately linear for smaller energies, and saturate near to the normal state values for higher energies.

It is tempting to assign the evolution, \textit{K vs. E}, to ``instantaneous'' sample heating, such that the spectra are recorded while $T>T_c(B_0)$. Indeed, eddy currents resulting from the high amplitude RF pulses provide a mechanism for absorption. Moreover, the heat capacity vanishes continuously in the limit $T\to0$, so the \textit{relative} temperature increase following a pulse at base temperature 1 K is many times greater than at, for example, 2K. Note that a check of the nuclear spin-lattice relaxation time $T_1$ is usually insensitive to such an effect, since the time scales for $T_1$ and electronic thermal relaxation are so different, $T_1>>\tau_{th}$. Therefore, in principle, it is possible that the spectra recorded following high energy pulses correspond to those of the normal state, while $T_1$ results would correspond to the superconducting state. For insight into the thermal conditions imposed by the RF pulses, time-synchronous measurements of the tank circuit reflected power were carried out. 
A summary of the conditions is as follows: an RF pulse (or sequence), as used for the NMR excitation, ``pumps'' the system. It is followed by a low-power RF probe, and the reflection is phase-sensitively detected using the NMR receiver. In this way we study the temporal changes to the reflected power, which depends on the sample response to the RF. Note that this is equivalent to an ac susceptibility measurement, and relates to RF shielding. Comparisons to the normal state are possible by measuring also the field dependence, including $B_0>B_{c2}$, or by warming to $T>T_c(B_0=0)$. Measurements were made at base temperature $T_{MC}=50$~mK.

\begin{figure}
\centering
\includegraphics[width=0.4\columnwidth]{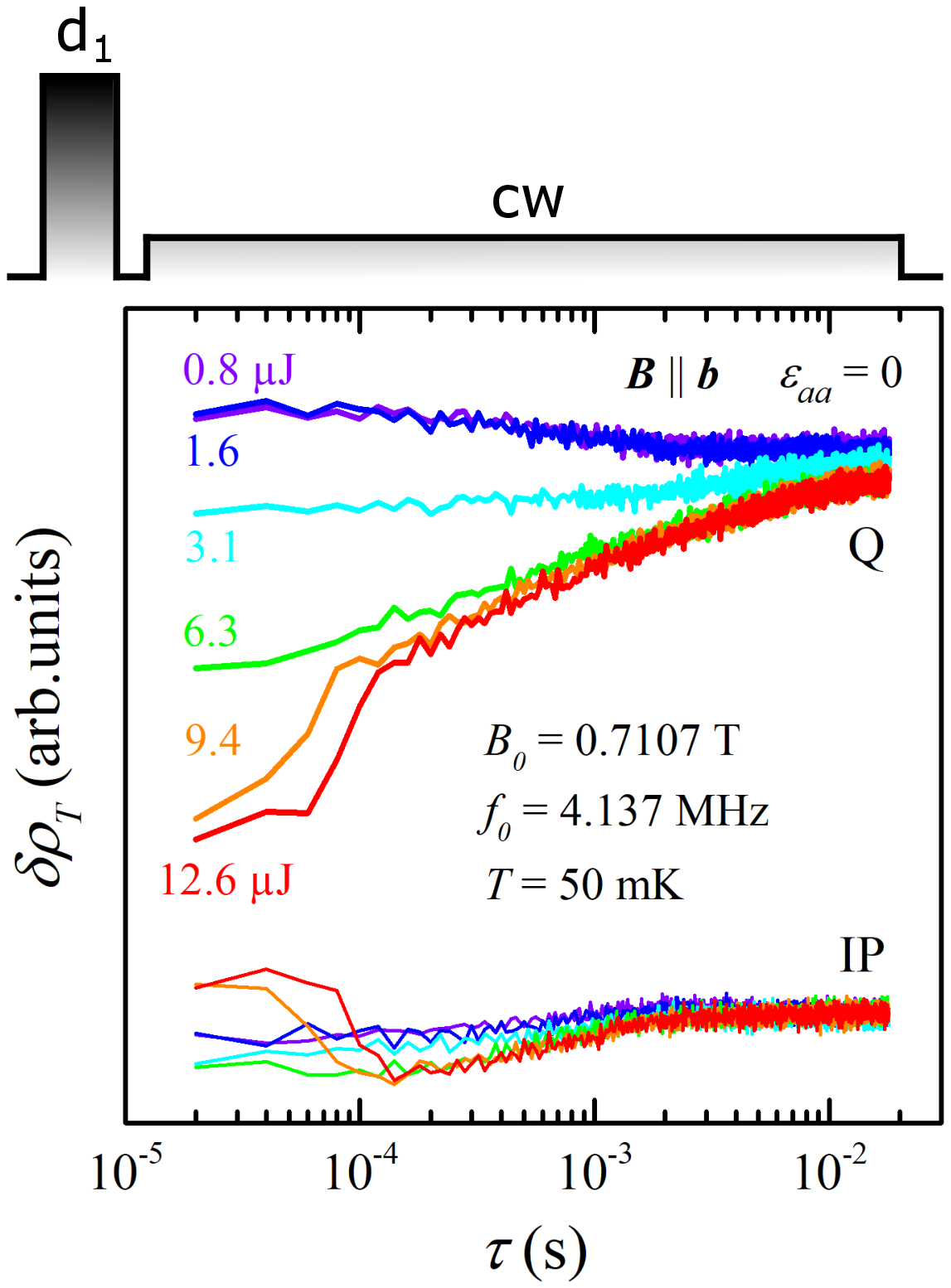}
\caption{\textbf{$|$ Transient effects following RF pulses, $\varepsilon_{aa}$=0.}
A short RF pulse of duration $d_1$ is applied at time $\tau=0$, followed by a low-power time-resolved cw measurement of the NMR tank circuit phase-sensitive reflected power, which is an RF equivalent to a complex ac susceptibility measurement. The data are presented as $\delta\rho_T$ vs. $E$, with $\delta\rho_T$ the changes to the reflection coefficient, and $E$ the energy of the pulse. Both in-phase (IP) and quadrature (Q) parts of $\delta\rho_T$ are strongly impacted, at short times $\tau$, for larger energies $E$. No similar time-dependence is observed when the sample is initially in the normal state. }
\label{fig:pulse-energy-transient}
\end{figure}

Our results for in-phase (IP) and quadrature (Q) components of the reflected power are shown in Fig.~\ref{fig:pulse-energy-transient}, where the applied pulse energies cover the range used for the NMR measurements. Note that the overall phase is arbitrary; otherwise the measurement is an RF equivalent to an ac susceptibility measurement, and therefore sensitive to screening currents in the sample. For sufficiently high energy pulses, the recovery to a steady state takes place via a two-step process. Aided by comparing to similar measurements carried out in varying fields (see SI), our interpretation is that the sample under study is initially responding as though it is in the normal state. This lasts for a period of about 100$\mu$s. A second, longer period of relaxation ($O$(1 ms)), occurs within the superconducting state. The longer time is likely due to changing vortex structure, motion and creep. No such time dependence is observed if the sample is initially in the normal state.  

In summarizing the results presented so far, the experiments indicate a reduced spin polarization upon entering the superconducting state at zero strain, as well as at $\varepsilon_v$. The field strengths relative to the upper critical field were $B_0/B_{c2}=0.55$ and 0.45, respectively. An important question is whether there is any indication for a change in order parameter symmetry between these limiting cases. Toward that end, spectra from normal and superconducting states were recorded for strains covering the interval $\varepsilon_{aa}=[0,\varepsilon_v]$; these are presented in the Supplementary Information. Collected at fields 0.7107 T, 1.1573 T, the data in the superconducting state vary continuously, with no discernible jump in $M_s$. Since one route to a symmetry change is via a first order phase transition, these results, when combined with the smooth variations of $B_{c2}$ (Fig. \ref{Hc2}) and the previously measured $T_c$\cite{Steppke-Science2017}, suggest this possibility unlikely.

\textbf{Discussion}

The key experimental finding that we report in this paper is our deduction that, at all applied uniaxial pressures, the spin susceptibility deduced from our Knight shift measurements is substantially suppressed at 20 mK from the normal state value. For unstrained samples, this conclusion differs from that previously drawn in the literature, and is therefore completely inconsistent with the $\mathbf{d} = \hat{z}(k_x \pm ik_y)$ order parameter, or indeed any odd-parity state with an out-of-plane $\mathbf{d}$.  Since such order parameters have been widely postulated to be the ones relevant to \sro\ for over twenty years, we believe that this represents a major advance in our understanding of this exemplar of unconventional superconductivity.

It is also important to note that although we can rule out specific odd-parity order parameters of the type described above, we cannot rule out all odd-parity states on the basis of our NMR data alone, because the uncertainty in estimating the quasiparticle background signal at $B_0/B_{c2}\sim0.5$ means that we cannot definitively distinguish odd-parity order parameters with in-plane $\mathbf{d}$-vectors (such as the A$_{1,2u}$ and B$_{1,2u}$ states in Table \ref{tab:OP}) from even parity states
at all measured strains.  By reducing the measurement field to 0.7107 T, we obtain a drop of 75\% in the Knight shift at $\varepsilon_{aa}=\varepsilon_v$, see Fig.~S5.  If that is all attributable to the spin susceptibility it would indeed rule out all the triplet states listed in Table~\ref{tab:OP}.  For the unstrained samples we observe a maximum Knight shift reduction of approximately 50\%, which would still be consistent with an A$_{1,2u}$ or B$_{1,2u}$ state.  We believe that further work on larger and more completely $^{17}$O enriched samples will enable us to determine whether the true reduction is more than 50\% in the unstrained case.  
We note that reconciling with the independent observations of time reversal symmetry breaking would require some amount of fine-tuning or some unusual physics.  For instance, one may consider a situation in which, accidentally, the A$_{1u}$ and B$_{1u}$ states had nearly identical transition temperatures.  In that case, one may imagine forming distinct domains, some stabilizing the A$_{1u}$ state, others the B$_{1u}$ state, and with non-trivial relative phases between such domains resulting in the breaking of time-reversal symmetry.  Since NMR is a local probe, domains of this kind would generally produce distinct line shapes which we do not observe.  In the even-parity sector, the tetragonal crystal field prevents in-plane paired states such as $d_{x^2-y^2}$ and $d_{xy}$ from having the same $T_c$, so the only plausible time reversal symmetry breaking order parameter would be of the form $d_{xz}+id_{yz}$, an exotic state for which the Cooper pairing would be between electrons in adjacent planes.  For any bulk time reversal symmetry breaking state, however, transition splitting under uniaxial pressure is expected, something that has not been observed either in our work or that of other groups.  Extending studies of time reversal breaking to strained \sro\ is therefore a priority for future work.  Overall, these are exciting times for research on \sro, with the promise of a fundamentally new framework with which to understand its enigmatic superconductivity.

\begin{table}
\centering
\begin{tabular}{|c|c|c|c|}
\hline
represent. & basis func. & Nodes & TRSB\\
\hline\hline
A$_{1u}$ & $\vec{d}=\hat{x}k_x+\hat{y}k_y$ & No & No\\
\hline
B$_{1u}$ & $\vec{d}=\hat{x}k_x-\hat{y}k_y$ & No & No\\
\hline
A$_{2u}$ & $\vec{d}=\hat{x}k_y-\hat{y}k_x$ & No & No\\
\hline
B$_{2u}$ & $\vec{d}=\hat{x}k_y+\hat{y}k_x$ & No & No\\
\hline
B$_{1g}$ & $\psi_d=k_x^2-k_y^2$ & vertical & No\\
\hline
B$_{2g}$ & $\psi_d=k_xk_y$ & vertical & No \\
\hline
E$_u$ & $\vec{d}=\hat{z}\left(k_x\pm ik_y \right)$ & No & Yes\\
\hline
E$_u$ & $\vec{d}=k_z\left(\hat{x}\pm i\hat{y}\right)$ & Horizontal & Yes\\
\hline
E$_{g}$ & $\psi_{chiral}=k_z(k_x\pm ik_y)$ & horizontal & Yes\\
\hline
\end{tabular}\caption{Irreducible representations for selected possible allowed $p-$ and $d-$ wave order parameters. The two E$_u$ states above belong to the same irreducible representation and can therefore coexist.  Hence, the horizontal nodes of the E$_u$ state with in-plane d-vector are not protected. }
\label{tab:OP}
\end{table}

\newpage

\begin{methods}

\subsection{Experimental.}

High-quality single crystalline Sr$_2$RuO$_4$ used for these measurements was grown by the floating-zone method as described elsewhere~\cite{Maeno-Nature1994}. Smaller pieces were cut and polished along crystallographic axes with typical dimensions 3$\times$0.3$\times$0.15 mm$^3$, with the longest dimension aligned with the $\mathbf{a}$-axis. $^{17}$O isotope ($^{17}I=5/2$, gyromagnetic ratio $^{17}\gamma_n$=$-$5.7719 MHz/T \cite{Hoult-2007}) spin-labelling was achieved by annealing in 50\% $^{17}$O-enriched oxygen atmosphere at 1050 $^\circ$C for 2 weeks \cite{Ishida-Nature1998}. The sample quality after annealing was confirmed by specific heat measurements which show $T_c\approx$1.44 K, essentially the same as before annealing \cite{Luo2018}. A piezoelectric type strain cell (Razorbill, UK) was employed to generate the uniaxial stress, and corresponding strain distortions along the $\mathbf{a}$-axis. Sample mounting between clamping plates included black Stycast 2850 (Loctite), with an effective compressive length about 0.9 mm. The strain values $\varepsilon_{aa}$ were estimated by a pre-calibrated capacitive dilatometer, and could reflect considerable systematic overestimation. Previous estimates place $\varepsilon_v\simeq-0.6\%$. For NMR  measurements, a small coil of $\sim$23 turns is made around the sample with 25 $\mu$m Cu wire, see Fig. 1c. NMR measurements were performed using a standard Hahn echo sequence with external magnetic field parallel with $\mathbf{b}$-axis, with strength determined by $^3$He nuclear resonance. Two samples in total were measured in this work, denoted as S1 and S2 respectively. S1 was measured at fixed strain $\varepsilon_v$ ($T_c=T_c^{max}$=3.5 K) and carrier frequency $f_0$=11.54 MHz ($B$=1.9980 T) at temperatures spanning both sides $T_c$. Two sets of measurements on S2 were carried out. In the first, a fixed carrier frequency $f_0$=6.7 MHz ($B$=1.1573 T) was chosen, for measurements at temperatures 25 mK (SC state) and 4.3 K (normal state), as a function of strain up to $-$0.58\%. The second set of measurements was a detailed study at zero strain. All measurements on S2 were performed using a dilution refrigerator (Oxford Kelvinox, UK) with the entire strain jig-with \sro\ crystal immersed inside the mixing chamber.

\subsection{NMR shift correction due to quadrupolar splitting.}

The NMR Knight shift, $K$, is generically defined as the percentage of the shift of resonance frequency with respect to a reference frequency $f_{ref}$=$\gamma_n B_0$, viz. $f$=$\gamma_n B (1+K)$. However, an additional field-dependent correction is necessary for nuclei with $I>$1/2, due to quadrupolar coupling to the electric field gradient. On a relative scale, the angle-dependent correction to the central transition is more important in weaker fields. It was numerically evaluated by diagonalizing the nuclear spin Hamitonian
\begin{equation}
H_{tot}=H_{Z}+H_{Q},
\label{Eq.1}
\end{equation}
where
\begin{equation}
H_{Z}=h \gamma_n (1+K) \mathbf{B}\cdot\mathbf{\hat{I}}
\label{Eq.2}
\end{equation}
characterizes the Zeeman effect, and
\begin{equation}
H_Q=\frac{eQV_{zz}}{4I(2I-1)}[3\hat{I_z}^2-\mathbf{\hat{I}^2}+\eta(\hat{I_x}^2-\hat{I_y}^2)]
\label{Eq.3}
\end{equation}
is the quadrupolar term. Here $h$ is Planck's constant, $\mathbf{\hat{I}}$=$(\hat{I_x},\hat{I_y},\hat{I_z})$ is nuclear spin operator, $Q$ is nuclear quadrupole moment, and $\eta$=($V_{xx}$$-$$V_{yy}$)/$V_{zz}$ is the asymmetry parameter with $V_{xx}$, $V_{yy}$ and $V_{zz}$ being the components of the electric-field gradient (EFG) tensor.

For our magnetic field values used in this work, the calculated corrections for different $^{17}$O sites are listed as following (in unit of kHz):

\begin{table}
\vspace*{-0pt}
\begin{center}
\def\temptablewidth{1\columnwidth}
{\rule{\temptablewidth}{1pt}}
\begin{tabular*}{\temptablewidth}{@{\extracolsep{\fill}}ccccc}
 $B$ (T)   & $f_{ref}$ (MHz)    & O(1)$_{\parallel}$       & O(1$^{\prime}$)$_{\perp}$   &      O(2)$_{b}$          \\ \hline
  0.7107   & 4.0980             & 1.0                      &     57.5                    &       43.0                 \\
  1.1573   & 6.6798             & 0.8                      &     36.0                    &       26.7                 \\
  1.9980   & 11.5323            & 0.5                      &     20.4                    &       15.6            \\
\end{tabular*}
\caption{Quadrupolar corrections, in kHz, to central transition resonant frequencies $f=^{17}\gamma B_0$, for the three oxygen sites and field strengths applied along the $b$-axis. The listed values apply to the zero strain case, and we have used $^{17}\gamma$=5.772 MHz/T \cite{Luo2018}.}
{\rule{\temptablewidth}{1pt}}
\end{center}
\end{table}

\end{methods}


\bibliography{sroSCsb}

\begin{addendum}
 \item We thank M. Ikeda and S. Kivelson for insightful conversations. A. P. acknowledges support by the Alexander von Humboldt Foundation through the Feodor Lynen Fellowship. Y. L. acknowledges a support by the 1000 Youth Talents Plan of China. This work was supported in part by the National Science Foundation (DMR-1709304).  Work at Los Alamos was supported by the Los Alamos National Laboratory LDRD Program.
 \item[Author Contributions]  Y.L., A.P. and S.E.B. conceived and designed the experiments. D.S., N.K., F.J., C.W.H. and A.P.M. prepared the crystal. E.D.B. characterized the sample and performed the spin labeling. A.P. and Y.L. performed the NMR measurements, supported by Y.S.S. and A.C.. A.P., Y.L., C.W.H., A.P.M. and S.E.B. discussed the data, interpreted the results, and wrote the paper with input from all the authors. The time-synchronous reflection experiments were performed by S.E.B., A.P., A.C. A.P. and Y.L. contributed equally to this work.
 \item[Competing Interests] The authors declare that they have no
competing financial interests.
 \item[Correspondence] Correspondence and requests for materials should be addressed to A. Pustogow (pustogow@physics.ucla.edu), Y. Luo (mpzslyk@gmail.com), or S. E. Brown (brown@physics.ucla.edu).
\end{addendum}

\FloatBarrier

\newpage

\setcounter{table}{0}
\setcounter{figure}{0}
\setcounter{equation}{0}
\renewcommand{\thefigure}{S\arabic{figure}}
\renewcommand{\thetable}{S\arabic{table}}
\renewcommand{\theequation}{S\arabic{equation}}

\begin{center}
{\bf \large
{\it Supplementary Information}\\
Pronounced drop of $^{17}$O NMR Knight shift in superconducting state of Sr$_2$RuO$_4$
}
\end{center}

\small
\begin{center}
A. Pustogow$^{1*}$, Yongkang Luo$^{1,2*}$, A. Chronister$^{1}$, Y.-S. Su$^{1}$, D. Sokolov$^{3}$, F. Jerzembeck$^3$, A. P. Mackenzie$^{3}$, C. W. Hicks$^3$, N. Kikugawa$^{4}$, S. Raghu$^5$, E. D. Bauer$^6$, and S. E. Brown$^1$\\

$^1${\it Department of Physics $\&$ Astronomy, UCLA, Los Angeles, CA 90095, USA;}\\
$^2${\it Wuhan National High Magnetic Field Center and School of Physics, Huazhong University of Science and Technology, Wuhan 430074, China;}\\
$^3${\it Max Planck Institute for Chemical Physics of Solids, Dresden 01187, Germany;}\\
$^4${\it National Institute for Materials Science, Tsukuba, Japan;}\\
$^5${\it Geballe Laboratory for Advanced Materials, Stanford University, Stanford, CA, 94305, USA;}\\
$^6${\it Los Alamos National Laboratory, Los Alamos, New Mexico 87545, USA.}\\

\date{\today}
\end{center}
\normalsize

\section{Orbitals and apparatus}

\begin{figure}[t]
\centering
\includegraphics[width=\columnwidth]{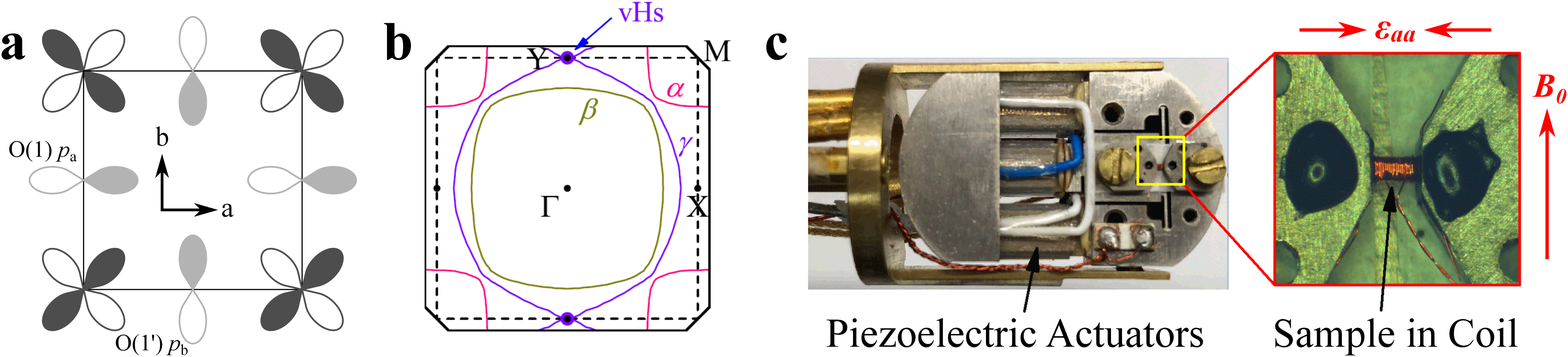}
\caption{\textbf{$|$ RuO$_2$ plane, with $d_{xy}-p$ hybridizing orbitals and experimental setup.} \textbf{a,} Depiction of Ru $d_{xy}$- and hybridizing O $p$-orbitals at the Y-point, which dominate formation of the $\gamma$ band. NMR shifts are measured at the O(1) and O(1$^{\prime}$) sites. \textbf{b,} Compressive $a$-axis stress shifts the $\gamma$ band Fermi surface to the zone boundary at Y.
\textbf{c,} Image of the strain device. The enlarged view highlights the \sro\ single crystal mounted between piezoelectric actuators, with $B_0$ parallel to the $b$-axis and $a$-axis compressive stress, $\varepsilon_{aa}$. The NMR coil covers the free part of $\approx 1$~mm length.}
\label{OrbitalsApparatus}
\end{figure}

The geometry of the experiment is shown in Fig. \ref{OrbitalsApparatus}. The crystal structure consists of layers of corner sharing O octahedra with Ru ions at the center of each. In panel \textbf{a}, we illustrate the planar coordination (at the Y point of the Brillouin Zone) of hybridizing Ru and O orbitals predominating the quasi-2D $\gamma$ band character. An in-plane magnetic field $B_0\parallel b$ yields inequivalent O(1) and O(1$^{\prime}$) sites with distinct Knight shifts. Uniaxial deformation along the $a$-axis pushes the $\gamma$ states at $E_F$ towards the Brillouin zone boundary (Fig.~\ref{OrbitalsApparatus}b) giving rise to a van Hove singularity in the density of states~\cite{Steppke-Science2017,Luo2018}. In Fig.~\ref{OrbitalsApparatus}c we show the sample mounted in the strain cell where $\epsilon_{aa}\parallel a$ and $B_0\parallel b$. The NMR coil was wrapped around the free part of the crystal, thus covering only the non-glued area that is subject to uniaxial stress. More information on strain-dependent NMR experiments on \sro\ can be found in Ref.~\cite{Luo2018}.

\section{Upper critical field measurements and estimation of strain gradients}

\begin{figure}[ht]
\centering
\includegraphics[width=0.4\columnwidth]{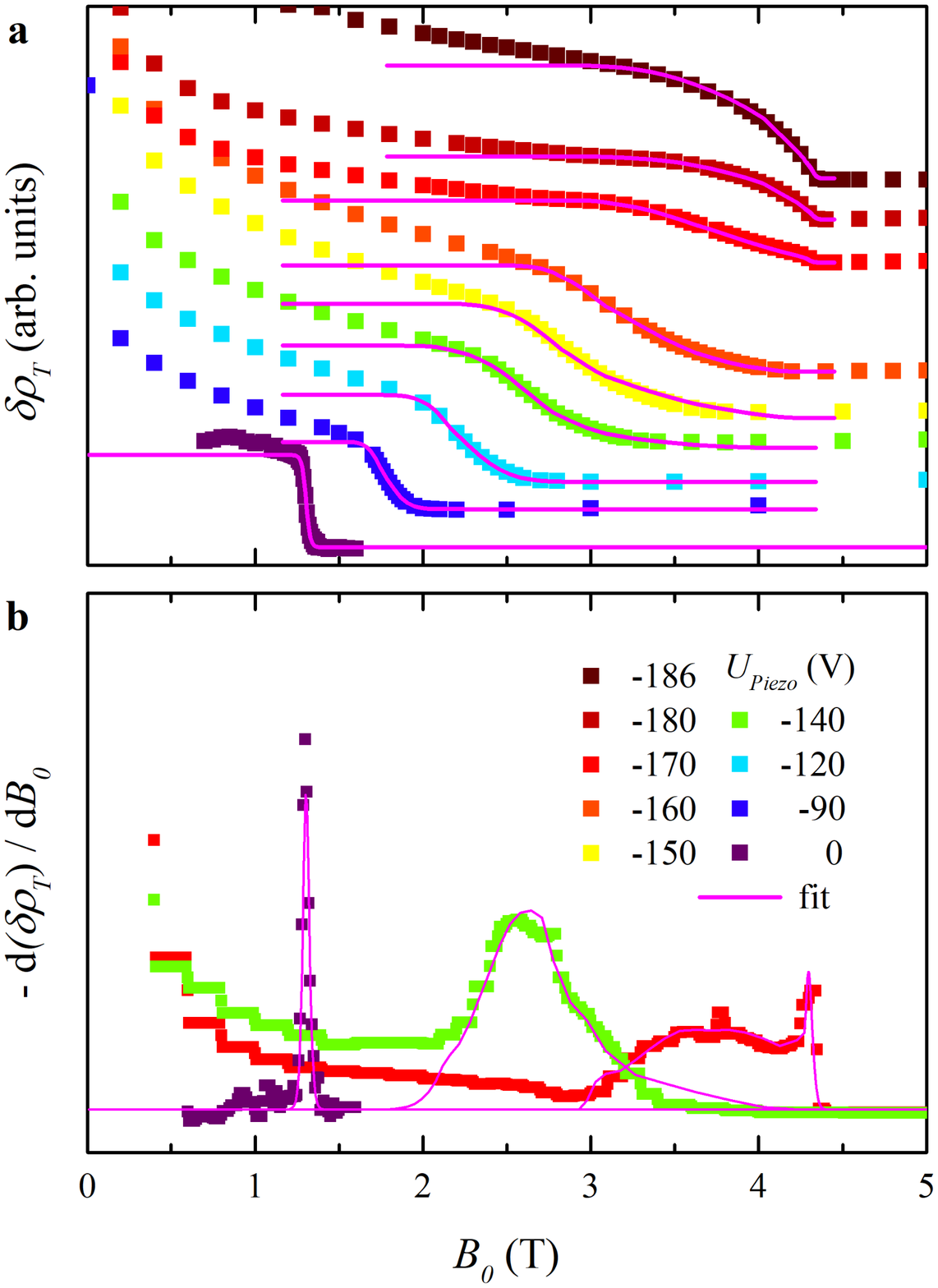}
\caption{\textbf{$|$ Estimation of strain gradients.} \textbf{a,} $B_{c2}$ in Fig.~1 was determined from magnetic field sweeps of the tank circuit reflected power. The broadening of the superconducting transition was modeled by a Gaussian strain distribution of half-width $\delta\varepsilon/\varepsilon_{aa}\simeq10\%$ (pink lines). \textbf{b,} The fitting curves also match with the corresponding derivative $d(\delta\rho_T)/dB_0$. For clarity, only a subset of the measured fields is shown.
}
\label{strain-gradient}
\end{figure}

As summarized in Fig.~1, the upper critical field $B_{c2}$ was determined at base temperature $T_{MC}=20$ mK by measuring the field dependence of the power reflected from the NMR tank circuit. More specifically, the frequency is set close to the tune/match condition. Variations in the reflection coefficient $\delta\rho_T$ relate to changes in the complex load impedance. Consequently, the measurement is equivalent to an ac susceptibility experiment, and is sensitive to screening current changes that occur, for example, when the system is driven from the superconducting to the normal state by the magnetic field. $B_{c2}$ is taken as the steepest slope at the transition midpoint, \textit{i.e.}, the maximum of $d(\delta\rho_T)/dB$, as plotted in Fig.~\ref{strain-gradient}a,b for the respective strain potential bias, $U_{Piezo}$. The 'onset' and 'lower end' values were defined as the kinks in the derivative $d(\delta\rho_T)/dB$ above and below $B_{c2}$, respectively. The superconducting transition exhibits considerable broadening with increasing compressive strain $\varepsilon_{aa}$. To model the apparently smeared transition, we assumed a Gaussian distribution of strains, and used the resulting distribution in $B_{c2}$ to generate the solid magenta curves in Fig.~\ref{strain-gradient}. The curves shown correspond to strain variations of 10\% of the normalized value, $\varepsilon_{aa}/\varepsilon_v$. The approach provided a self-consistent method for estimating the relative importance of strain distributions, and suitably describes our observations.

\newpage

\section{Assessing the Zero-Strain Position}

\begin{figure}[ht]
\centering
\includegraphics[width=0.7\columnwidth]{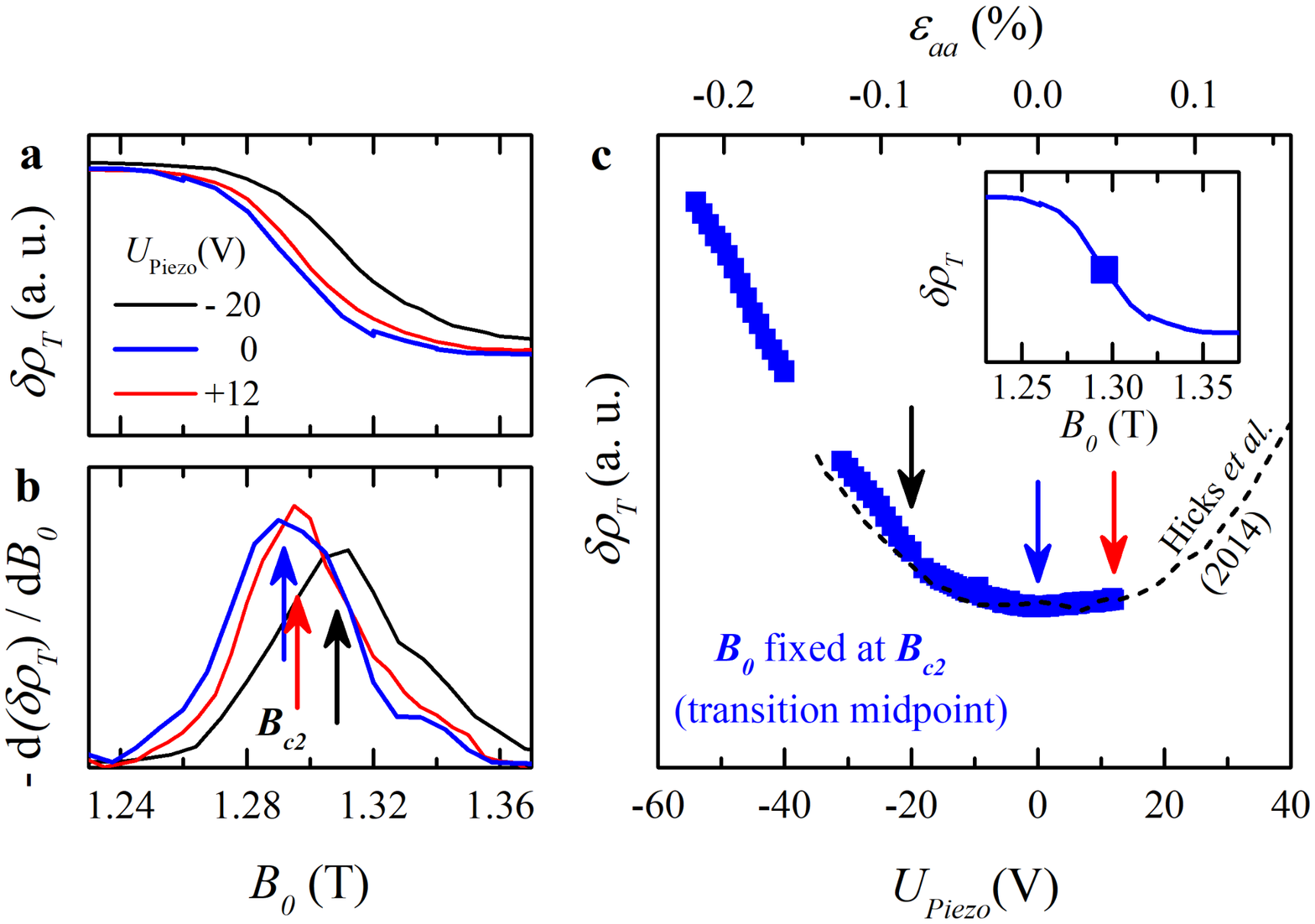}
\caption{\textbf{$|$ Tuning the crystal to zero strain.} \textbf{a,} Measurements of the reflected power $\delta\rho_T$ at low strain indicate that $B_{c2}$ has a minimum near $U_{Piezo}=0$~V. \textbf{b,} The derivative $d(\delta\rho_T)/dB_0$ illustrates that $B_{c2}$ associated with the largest rate of change at the transition midpoint first decreases when reducing compression ($U_{Piezo}=-20\rightarrow 0$~V), followed by a slight increase upon tensile strain ($U_{Piezo}=0\rightarrow +12$~V). \textbf{c,} The coil impedance was measured at the transition midpoint ($B_0$ fixed at $B_{c2}$), providing a sensitive measure of modifications upon changing strain. In intervals of 20~V or less, $B_{c2}$ was determined by a field sweep and afterwards $B_0$ was set to the new transition midpoint. The results (solid blue squares) were corrected for the different $B_0$ yielding one half of a parabola centered around [-10~V;10~V], very similar to the strain dependence of $T_c$~\cite{Hicks2014} (dashed black line).}
\label{zero-strain}
\end{figure}

The ``zero-strain'' condition attributed to the spectra shown in Fig.~3 was determined by taking the minimum in $B_{c2}$ as a proxy. Note that since there is differential thermal contraction between strain device and sample, the strain-free condition needed to be assessed \textit{in situ}. The determination was carried out in two steps. First, for a range of discrete values of piezo bias $U_{piezo}$ ranging to positive and negative values about 0 V, $B_{c2}(U_{piezo})$ was determined using field sweeps and recording the reflected power, just as for the measurements described in Fig.~1, and Fig.~\ref{strain-gradient}. Example sweeps are shown in Fig.~\ref{zero-strain}a,b, from which the minimum was found to be near $U_{piezo}=0$ V. A more accurate determination was made by first setting the initial field to $B_{c2}(U_{piezo}=0$~V. That is, the conditions were set to the transition midpoint shown in Fig.~\ref{zero-strain}c. Then, changes in reflected power were recorded on sweeping $U_{piezo}$ about zero. Here again, we found the zero-strain condition indistinguishable from $U_{piezo}$=0 V. Thus, our experiments at low pulse energy indeed probe the superconducting properties of \sro\ at zero strain.

\newpage

\section{Time-synchronous reflected power response}
\begin{figure}[hb]
\centering
\includegraphics[width=0.6\columnwidth]{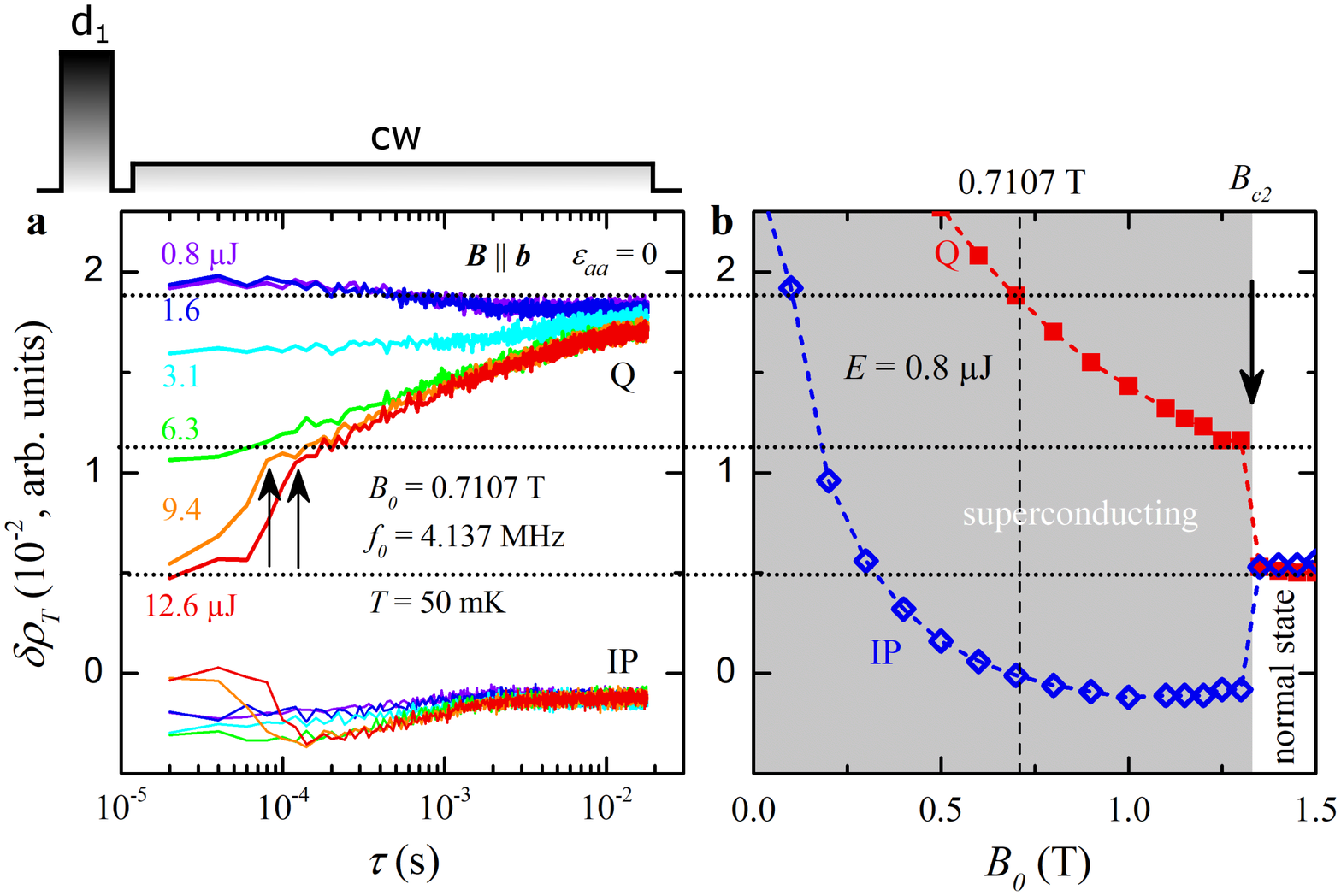}
\caption{\textbf{Transient effects associated with normal-state response. a}, The transient components of reflected power are plotted as a function of pulse energy $E$ and time $\tau$ after the pulse, cf. Fig.~4. \textbf{b}, The magnetic field dependence of reflected power was recorded at $E = 0.8$~$\mu$J. The changes in $\delta\rho_T(B_0)$ on increasing $B_0$ from the measurement field (0.7107 T) to $B_0>B_{c2}$ match well with the time-dependent recovery in \textbf{a}, as indicated by the horizontal dotted lines. Both channels (IP and Q) document a variation in $\delta\rho_T$ that results from a transition between normal and superconducting states around $\tau\approx100$~$\mu$s.
}
\label{ReflectionsNormalSC}
\end{figure}

Our interpretation of the time-synchronous reflected power measurements were aided by comparable measurements carried out in variable field conditions, with the goal to contrast normal and superconducting state responses. Results are shown in Fig.~\ref{ReflectionsNormalSC}, where the applied pulse energies cover the range used for the NMR measurements. First consider Fig. S4b, which documents a measurement of the power reflected from the NMR tank circuit as the magnetic field is varied to a strength exceeding $B_{c2}$. The jump at 1.3 T corresponds to the transition to the normal state. The evolution at lower fields is presumably associated with the response of a changing vortex structure, including density and characteristic length scales. The vertical dashed line is the measurement field for the spectra shown in Fig.~3, $B_0=0.7107$ T. Both components of $\delta\rho_T$, in-phase (IP) and quadrature (Q), show a pronounced time dependence for large pulse energies. Our interpretation is that the sample under study is initially responding as though it is in the normal state. The relaxation back to the static superconducting state appropriate for the applied field occurs via a two-step relaxation process. First it is normal for $\tau\leq100$~$\mu$s, followed by a slower relaxation on the order of a few milliseconds, while in the superconducting state. The horizontal dotted lines in Fig.~\ref{ReflectionsNormalSC} indicate the values $\delta\rho_T(B_0)$ right above and below the superconducting transition, as well as the static state at 0.7107~T. The longer time is likely due to changing vortex structure, motion and creep. No such time dependence is observed if the sample is initially in the normal state.

\section{Strain dependent NMR shifts}
\begin{figure}[ht]
\centering
\includegraphics[width=.4\columnwidth]{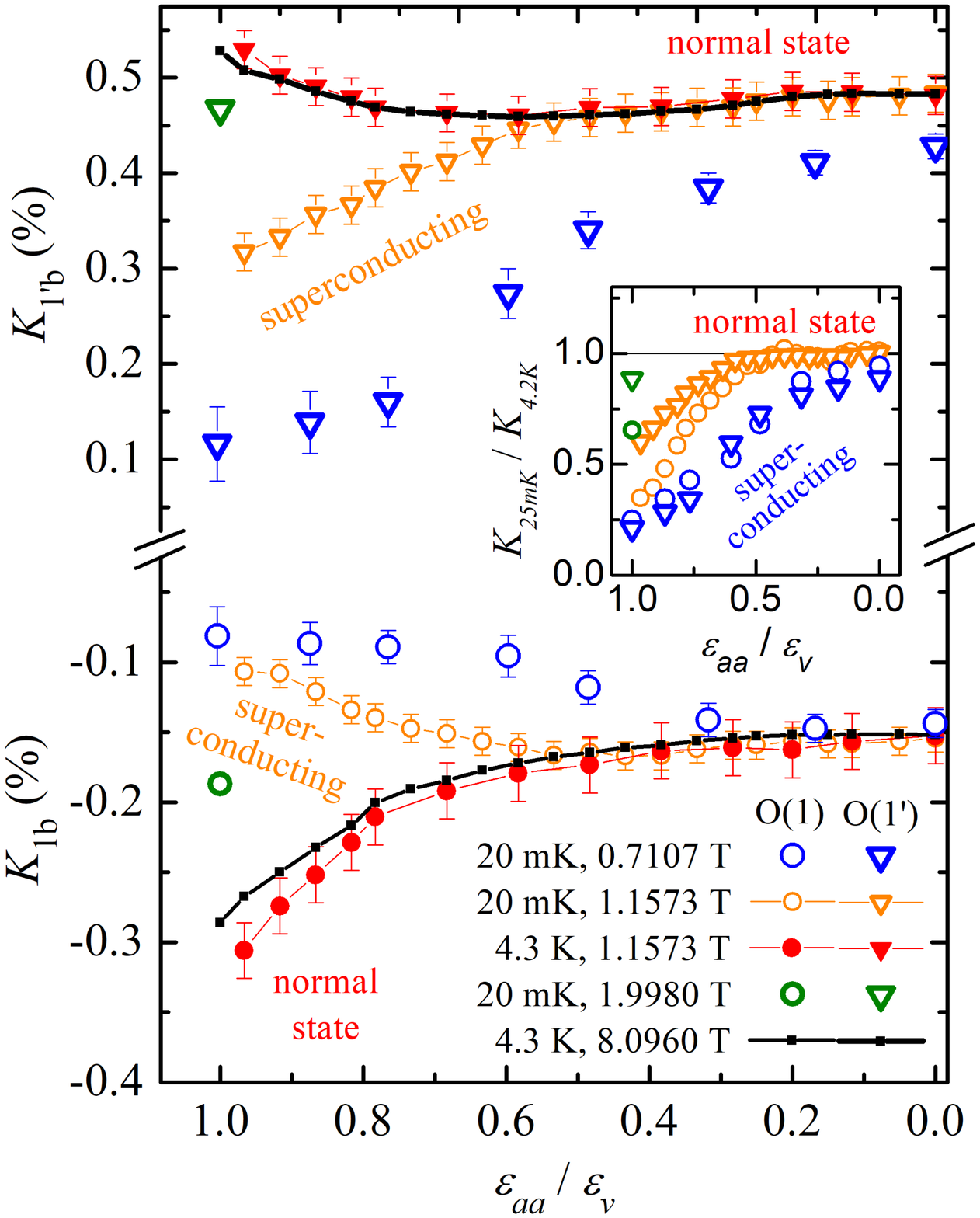}
\caption{\textbf{$|$ Strain dependence of $^{17}$O Knight shifts in superconducting and normal states.} Contrast of shifts in normal and superconducting states for strains covering the range $\varepsilon_{aa}=[0,\varepsilon_v]$. The top and bottom parts of the panel show the O1$^{\prime}$ and O1 sites, respectively. Normal state is indicated by solid symbols (black and red) recorded at 4.3 K and two different field strengths~\cite{Luo2018}. Open symbols correspond to an equilibrium temperature of 20 mK, hence within the superconducting state for sufficiently high $T_c$ realized by large strain and small magnetic field. Blue and orange symbols correspond to field strengths 0.7107 T and 1.1573 T, respectively. The results from $B_0=1.9980$ T are shown in green ($\varepsilon_{aa}=\varepsilon_v$; cf. Fig.~2).}
\label{Shift-Strain}
\end{figure}

The results from strain-dependent studies of the NMR Knight shifts are shown in Fig.~\ref{Shift-Strain}, with data from the superconducting state depicted by open symbols (equilibrium temperature 20 mK), and from the normal state indicated by solid symbols (equilibrium temperature 4.3 K)~\cite{Luo2018}. In addition to the results from Fig.~2 (green; $T=20$~mK, $B_0=1.9980$~T), two different fields are shown for each temperature. The suppression of Knight shifts $K$ in the superconducting state reaches about 80\% for strain near $\varepsilon_v$ in the case of $B_0$=0.7107~T (blue). The effect of superconductivity weakens on lowering the strain, as $T_c$ and $B_{c2}$ both decrease. Since the spectra are generated by the standard echo sequence for these data, the pulse energies are large -- about $E\approx 10$~$\mu$J, which is comparable to the top trace in Fig.~3 of the main text. Thus, the results at low and zero strain correspond to the normal-state shifts, which is very obvious in the 20~mK results at 1.1573~T (orange) that deviate from the 4.3~K data only for $\varepsilon_{aa}>\varepsilon_v/2$. While the reduction of $K$ is generally more pronounced for 0.7107~T, also these data approach the normal-state values towards $\varepsilon_{aa}\rightarrow 0$. Evidently, the impact of the pulse energy decreases when the sample is strained because $T_c$ increases and the heating effect is not sufficiently strong to drive the crystal to the normal state. The results appear to vary continuously with applied field and strain, hence they provide no indication for a first-order phase transition between different superconducting order parameter symmetries.

\end{document}